\documentclass[10pt,english,journal]{IEEEtran}
\usepackage{geometry}
\usepackage{amsmath}
\usepackage{xpatch}
\usepackage{amssymb}
\usepackage{esint}
\usepackage{mathtools}
\usepackage{amsthm}
\usepackage{nicefrac}
\usepackage{bigints}
\usepackage{blkarray, bigstrut}
\usepackage{physics}
\usepackage{calligra}
\usepackage{graphicx}
\usepackage{dsfont}
\usepackage{array,ragged2e}
\usepackage{enumitem}
\usepackage{etoolbox}
\usepackage{babel}
\usepackage{lipsum}
\usepackage{psfrag}
\usepackage[ruled,lined,linesnumbered]{algorithm2e}
\usepackage{algorithmic}
\usepackage{balance}
\usepackage{float}
\usepackage{hyperref}
\usepackage{xcolor}
\usepackage{subcaption}

\theoremstyle{definition}

\SetKw{KwBy}{by}

\makeatletter
\newcommand{\removelatexerror}{\let\@latex@error\@gobble}
\makeatother

\definecolor{lightgraybg}{RGB}{240, 240, 240}

\xpatchcmd{\proof}{\hskip\labelsep}{\hskip5\labelsep}{}{}
\makeatletter
\xpatchcmd{\proof}{\@addpunct{.}}{\@addpunct{:}}{}{}
\makeatother

\renewcommand\[{\begin{equation}}
\renewcommand\]{\end{equation}} 
\usepackage{listings}
\usepackage{fancyvrb}
\usepackage{framed}
\usepackage{courier}

\definecolor{dkgreen}{rgb}{0,0.3,0}
\definecolor{gray}{rgb}{0.5,0.5,0.5}
\definecolor{codegreen}{rgb}{0,0.6,0}
\definecolor{codegray}{rgb}{0.5,0.5,0.5}
\definecolor{codepurple}{rgb}{0.58,0,0.82}
\definecolor{backcolour}{rgb}{0.95,0.95,0.92}
\definecolor{skyblue}{RGB}{135,206,235}

\lstdefinestyle{mystyle}{
    backgroundcolor=\color{backcolour}, 
    commentstyle=\color{codegreen}, 
    keywordstyle=\color{magenta}, 
    numberstyle=\tiny\color{codegray}, 
    stringstyle=\color{codepurple}, 
    basicstyle=\ttfamily\footnotesize, 
    breakatwhitespace=false, 
    breaklines=true, 
    captionpos=b, 
    keepspaces=true, 
    numbers=left, 
    numbersep=5pt, 
    showspaces=false, 
    showstringspaces=false, 
    showtabs=false, 
    tabsize=2 
}
\makeatletter
\newcommand*{\rom}[1]{\expandafter\@slowromancap\romannumeral #1@}
\makeatother

\usepackage{siunitx}
\usepackage{tabu}
\usepackage{booktabs}
\usepackage{multirow}
\usepackage{capt-of}
\usepackage{array}
\usepackage{arydshln}
\usepackage{cite}

\newcommand{\comment}[1]{}

\makeatletter
\patchcmd{\@maketitle}
  {\addvspace{0.5\baselineskip}\egroup}
  {\addvspace{-1\baselineskip}\egroup}
  {}
  {}
\makeatother

\begin{document}

\title{LLM-Based Agentic Negotiation for 6G: Addressing Uncertainty Neglect and Tail-Event Risk}

\author{
Hatim~Chergui,~\IEEEmembership{Senior~Member,~IEEE}, Farhad~Rezazadeh,~\IEEEmembership{Member,~IEEE}, Mehdi~Bennis,~\IEEEmembership{Fellow,~IEEE}, \\ Merouane~Debbah,~\IEEEmembership{Fellow,~IEEE},
and Christos~Verikoukis,~\IEEEmembership{Senior~Member,~IEEE}

\IEEEcompsocitemizethanks{\IEEEcompsocthanksitem H. Chergui is with the i2CAT Foundation, Spain. (e-mails: chergui@ieee.org, name.surname@i2cat.net)}
\IEEEcompsocitemizethanks{\IEEEcompsocthanksitem F. Rezazadeh is with the Technical University of Catalonia (UPC), Spain (e-mail: farhad.rezazadeh@upc.edu).}
\IEEEcompsocitemizethanks{\IEEEcompsocthanksitem M. Bennis is with the University of Oulu, Finland (e-mail: mehdi.bennis@oulu.fi).}
\IEEEcompsocitemizethanks{\IEEEcompsocthanksitem M. Debbah is with the Research Institute for Digital Future, Khalifa University, 127788 Abu Dhabi, UAE (e-mail: merouane.debbah@ku.ac.ae).}
\IEEEcompsocitemizethanks{\IEEEcompsocthanksitem C. Verikoukis is with ISI/ATH and University of Patras, Greece. (e-mail: cveri@isi.gr)}
}

\maketitle

\begin{abstract}
A critical barrier to the trustworthiness of sixth-generation (6G) agentic autonomous networks is the \emph{uncertainty neglect bias}; a cognitive tendency for large language model (LLM)-powered agents to make high-stakes decisions based on simple averages while ignoring the tail risk of extreme events. This paper proposes an unbiased, risk-aware framework for agentic negotiation, designed to ensure robust resource allocation in 6G network slicing. Specifically, agents leverage Digital Twins (DTs) to predict full latency distributions, which are then evaluated using a formal framework from extreme value theory, namely, Conditional Value-at-Risk (CVaR). This approach fundamentally shifts the agent's objective from reasoning over the mean to \emph{reasoning over the tail}, thereby building a statistically-grounded buffer against worst-case outcomes. Furthermore, our framework ensures full uncertainty awareness by requiring agents to quantify \emph{epistemic} uncertainty---confidence in their own DTs predictions---and propagate this meta-verification to make robust decisions, preventing them from acting on unreliable data. We validate this framework in a 6G inter-slice negotiation use-case between an eMBB and a URLLC agent across 200 trials. The results demonstrate the profound failure of the biased, mean-based baseline, which systematically violates the strict URLLC SLA 11 times. Our unbiased, CVaR-aware agent successfully mitigates this bias, eliminating SLA violations entirely and significantly reducing the 99.999th-percentile (p99.999) latencies by up to 51.7\%. We show this reliability comes at the rational and quantifiable cost of reduced energy savings, exposing the \emph{false economy} of the biased approach. Crucially, executing our framework with an \texttt{otel-llm-1b-it} model on a single NVIDIA RTX A4000 GPU achieves sub-1.5-second inference times, validating the feasibility for non-real-time RIC use-cases.
\end{abstract}


\section{Introduction}
\IEEEPARstart{T}{he} evolution towards 6G networks necessitates unprecedented levels of operational autonomy, targeting the TM Forum's Levels 4 (Closed-Loop Automation) and 5 (Full Autonomy) \cite{tmforum2021autonomous}. Achieving this paradigm shift requires moving towards \emph{agentic systems} \cite{ferrag2025llmreasoningautonomousai}, which must be capable of reasoning, planning, and negotiating at the goal level to dynamically manage network functions, slice orchestration, and service assurance in highly complex and volatile environments. As these LLM-powered agentic systems are deployed to manage critical 6G functions, ensuring their reliability and safety is paramount. It is crucial, however, to recognize the vulnerability of agentic systems to what are known as cognitive biases---systematic distortions that lead decisions away from fully rational judgment. Although this notion originates in human psychology, such biases can be transferred to artificial agents through their design, data, or learning processes, and subsequently reinforced over time. 

The foundational understanding of these systematic errors was established by Tversky and Kahneman in their seminal work, \emph{Judgment under Uncertainty: Heuristics and Biases} \cite{Tversky1974}, which demonstrated that human decision-making often relies on cognitive shortcuts---heuristics---that give rise to predictable and systematic deviations from rationality. These biases permeate every stage of an agent's behavior, influencing perception, reasoning, decision-making, and action execution. More recently, \cite{Xie2024} have emphasized the increasing relevance of such biases in artificial systems, particularly examining their emergence in large language models operating within multi-agent frameworks---an architecture of growing importance for decentralized 6G networks. The pervasive impact of bias can be traced across the entire agentic pipeline, from perception to action, and can be broadly categorized into four principal layers, namely,
i) \textbf{Data-level biases:} These originate from imbalances in training data, including historical inequities, cultural distortions, and sampling limitations. For example, a model trained predominantly on data from regions with legacy network infrastructure may exhibit a preference for outdated architectures when deployed in more advanced environments, resulting in suboptimal performance; ii) \textbf{Prompt-level biases:} The formulation of prompts can introduce framing effects that shape an agent's interpretation of a task. For instance, instructing a network agent to ``maximize throughput at all costs'' may bias it toward solutions that neglect critical considerations such as latency or fairness; iii) \textbf{Reasoning-level biases:} The internal decision-making processes of agents are susceptible to heuristic-driven distortions. An agent responsible for dynamic resource allocation may exhibit an availability bias, disproportionately prioritizing network slices associated with recent or frequent requests. Similarly, a security agent may display confirmation bias by favoring evidence that supports an existing threat model while overlooking novel attack patterns. These behaviors mirror human cognitive shortcuts, where easily accessible or pre-weighted information disproportionately influences outcomes; and iv) \textbf{Tool-use biases:} Biases also arise in how agents interact with external tools, including memory systems, data sources, and APIs. For example, recency or primacy effects may lead an agent to prioritize recently observed network logs over a more comprehensive historical dataset, resulting in short-sighted decisions. Additionally, authority bias may cause undue reliance on a single trusted data source or familiar API, even when more appropriate or diverse alternatives are available.

\subsection{Related Work}
\subsubsection{\textbf{Cognitive Biases in Agentic Systems}}
A comprehensive contribution in this domain is \cite{chergui2025tutorialcognitivebiasesagentic}, which presents a structured tutorial on well-known cognitive biases in 6G agentic systems, focusing on their definitions, mathematical formulation, emergence in 6G LLM-based agents and mitigation strategies at both the agent and 6G system levels, while providing practical use-cases with an accompanying source code. In~\cite{unmasking}, the authors demonstrate that iterative discussions amplify bias, forming conversational echo chambers as agents converge on consensus. This shows distortions can arise from interaction dynamics, not just pretrained knowledge. Structurally, the \emph{fairness in agentic AI} framework~\cite{fairness} examines how systemic distortions arise from decentralized collaboration, linking ethical alignment to negotiation constraints. Besides, the \emph{hidden profile benchmark}~\cite{assessing} shows that LLM agents often fail to share critical, unevenly distributed information, mirroring human groupthink and highlighting vulnerabilities in reasoning diversity. 

\begin{figure*}[htbp]
    \centering
\includegraphics[width=0.8\textwidth]{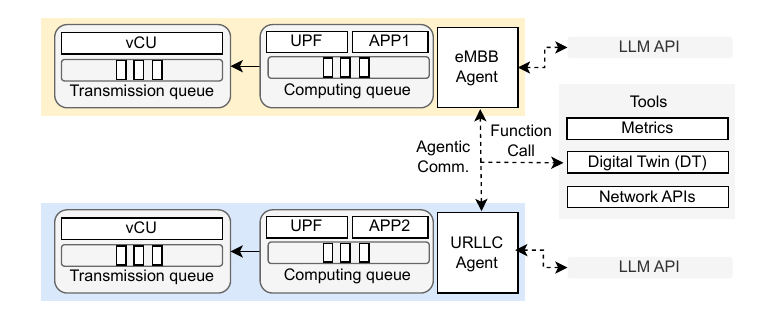}
    \caption{Agentic AI-driven 6G edge-RAN slicing.}
    \label{fig:ran_slicing}
\end{figure*}

\subsubsection{\textbf{Risk-Aware Optimization in 6G}}
Risk-based optimization has likewise been introduced in various 6G setups. For instance, \cite{risk1} introduces a risk-aware status-updating policy for real-time IoT monitoring---highly relevant for ultra-reliable, low-latency 6G systems---by jointly minimizing average age of information (AoI), tail AoI loss, and energy cost. Using CVaR to capture rare but severe AoI spikes, the authors reformulate an otherwise intractable history-dependent control problem as a standard MDP augmented with risk-level variables. Besides, \cite{risk2} presents a risk-sensitive resource allocation method for handling URLLC traffic while safeguarding ongoing eMBB transmissions. Using CVaR to quantify the risk of degrading low-rate eMBB users, the resulting approach efficiently allocates resources to incoming URLLC packets while simultaneously maintaining both eMBB protection and URLLC reliability.

\subsubsection{\textbf{Bayesian Digital Twins}}
In \cite{bay_dt1}, the authors highlight uncertainty quantification as a foundational requirement for reliable Digital Twins, reviewing how Bayesian and probabilistic methods are essential for capturing modeling errors, data noise, and operational variability. Besides, \cite{bay_dt2} advances this perspective by developing a Digital Twin based on a nonparametric Bayesian network that explicitly models and propagates epistemic uncertainty in complex-system degradation. Through real-time updating with Gaussian particle filtering and Dirichlet process mixture modeling, the DT continuously reduces uncertainty, adapts its structure, and delivers more accurate and trustworthy health monitoring. Finally, \cite{bay_dt} introduces a Bayesian Digital Twin framework for wireless systems that explicitly models epistemic uncertainty arising from limited or imperfect PT-to-DT data.

\subsection{Contributions}
While existing work identifies agentic biases, risk-based optimization, and Bayesian digital-twins, it overlooks a critical requirement for robust autonomy in 6G (TM Forum Levels 4/5): \emph{uncertainty quantification and communication} in LLM-based agentic systems. Autonomous components must convey the confidence of their decisions. Failure to model this introduces \emph{uncertainty neglect bias}, where low-confidence predictions are treated as high-confidence facts. This erroneous certainty can propagate throughout the multi-agent system, causing cascading failures, suboptimal resource allocation, and ultimately undermining network stability. This paper addresses this critical gap by:
\begin{itemize}
    \item Proposing a risk-aware negotiation framework for 6G LLM agents that mitigates uncertainty neglect bias by shifting agentic reasoning from simple averages to an \emph{Uncertainty-Adjusted Conditional Value-at-Risk} ($\widetilde{\text{CVaR}}_{\alpha}$). This mathematically enforces a strict Greedy Guardband to secure tail-end latency risk at a practical confidence level of $\alpha=0.99999$ for critical 6G services.
    \item Introducing a mechanism for epistemic uncertainty awareness that compels agents and internal tools (Digital Twins) to quantify and \emph{propagate} prediction confidence. This dynamically penalizes low-confidence data, preventing rigid decisions based on unreliable state estimations.
    \item Designing a hybrid, privacy-preserving architecture that leverages locally deployed, fast-inference 1-billion-parameter LLM \texttt{otel-llm-1b-it} \cite{otel2026}) for strategic reasoning. This is further augmented by a deterministic 2D Proportional-Integral-Derivative (PID) algorithm to guarantee multi-resource (Bandwidth and Edge CPU) constraint feasibility.
    \item Validating the framework via 200 trials of a 6G eMBB/URLLC inter-slice negotiation, demonstrating that the proposed approach strictly eliminates tail SLA violations, captures the trade-off between reliability and energy efficiency, and proves edge-deployable real-time capabilities via sub-1.5-second inference bounds on a single RTX A4000 GPU.
\end{itemize}

\section{Network Slicing Model and Problem Formulation}
\label{sec:system_model}

\subsection{Network Slicing Queuing Model and Digital Twin}
We consider a network slicing architecture spanning an Edge computing domain and a Radio Access Network (RAN) domain as depicted in Figure \ref{fig:ran_slicing}. Service requests for a slice $i$ first arrive at the Edge, incurring a computation latency $L^{\text{edge}}_i$. Processed packets are then enqueued for transmission over the wireless RAN, incurring a transmission latency $L^{\text{RAN}}_i$. The total end-to-end (E2E) latency is $L_i = L^{\text{edge}}_i + L^{\text{RAN}}_i$ \cite{unbiased}.

In our setup, both domains are resource-constrained. Agents must negotiate for a partition of the total available RAN bandwidth ($B_{\text{total}}$) and the total Edge CPU capacity ($F_{\text{total}}$). An agent $i$'s action is thus a vector $a_i = (b_i, f_i)$, where $b_i$ is its allocated bandwidth and $f_i$ is its allocated CPU. These actions determine the service capacity in each queue, and thus the latencies $L^{\text{edge}}_i$ and $L^{\text{RAN}}_i$. The agent's objective is to manage the total latency $L_i$ by controlling both $b_i$ and $f_i$.

Each agent $i$ is equipped with a private Digital Twin (DT) based on queuing theory. At each discrete time interval $t$ of duration $\tau$, a number of bits $\Lambda_{i,t}$ arrive at the Edge, governed by a time-varying stochastic process,
\begin{equation}
\Lambda_{i,t} = \lambda_{i,t} \cdot \tau,
\end{equation}
The computation queue at the edge, $Q^{(e)}_{i,t}$, evolves as,
\begin{equation}
Q^{(e)}_{i,t+1} = \max\left(0, Q^{(e)}_{i,t} - D^{(e)}_{i,t}\right) + \Lambda_{i,t},
\end{equation}
where $D^{(e)}_{i,t}$ is the number of bits processed by the Edge. This is determined by the agent's allocated CPU $f_i$ and a deterministic processing rate $C_{CPU}$,
\begin{equation}
D^{(e)}_{i,t} = \tau \times C_{i,t}^{(e)}(f_i) = \tau \cdot f_i \cdot C_{CPU}.
\label{eq: comp_lat}
\end{equation}
The RAN communication queue, $Q^{(r)}_{i,t}$, is updated based on the output of the compute queue,
\begin{equation}
\begin{split}
Q^{(r)}_{i,t+1} &= \max\left(0, Q^{(r)}_{i,t} - D^{(r)}_{i,t}\right) \\
&\quad + \min\left(Q^{(e)}_{i,t} + \Lambda_{i,t}, D^{(e)}_{i,t}\right),
\end{split}
\end{equation}
where $D^{(r)}_{i,t}$ is the number of bits transmitted. This is determined by the agent's action $b_i$ and the primary source of wireless uncertainty: the stochastic Spectral Efficiency (SE), $SE_{i,t}$, i.e.,
\begin{equation}
D^{(r)}_{i,t} = \tau \times C_{i,t}^{(r)}(b_i, SE_{i,t}) = \tau \cdot b_i \cdot SE_{i,t}.
\label{eq: radio_lat}
\end{equation}
Using Little's Law \cite{little}, the average E2E latency $L_{i,T}$ is the sum of the average queue lengths divided by the average arrival rate. The agent's objective is to control its action vector $a_i = (b_i, f_i)$ to keep $L_{i,T}$ below its SLA, while also minimizing a linear power consumption cost $P_i(a_i)$,
\begin{equation}
P_i(a_i) = P_{\text{static},i} + C_{BW} \cdot b_i + C_{CPU} \cdot f_i,
\end{equation}
where $C_{BW}$ and $C_{CPU}$ are bandwidth and CPU power costs, respectively.

\begin{figure*}[t]
\centering
\includegraphics[width=\textwidth]{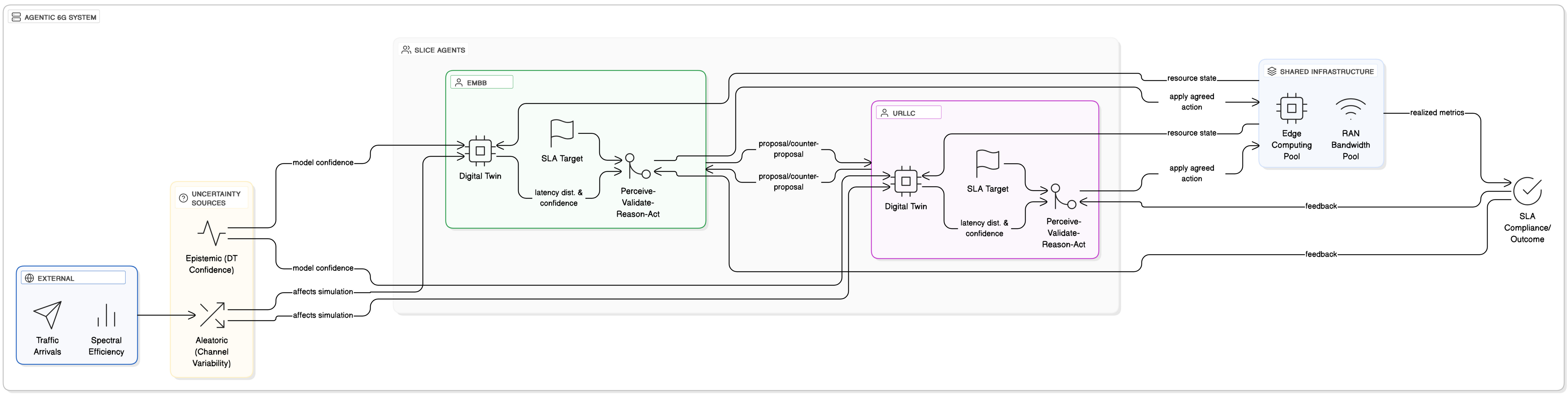}
\caption{Risk-aware agentic system concept.}
\label{fig: concept}
\end{figure*}

\subsection{Agentic Negotiation Framework}
For the sake of focusing on the effect of bias mitigation and without loss of generality, the system hosts two competing slices, $i \in \{1, 2\}$, each represented by an autonomous, LLM-based agent. These agents must negotiate to partition both the total available bandwidth $B_{\text{total}}$ and total edge CPU $F_{\text{total}}$. An agent's action is a vector $a_i = (b_i, f_i)$. The slices are: i) eMBB (Slice $i=1$), with an agent with a relaxed latency SLA (e.g., $L_{1, \text{SLA}} = 50\text{ms}$); and ii) URLLC (Slice $i=2$), with an agent with a strict latency SLA (e.g., $L_{2, \text{SLA}} = 10\text{ms}$). These two agents must continuously negotiate to find a mutually agreeable partition $(a_1, a_2)$ such that $b_1 + b_2 \le B_{\text{total}}$ and $f_1 + f_2 \le F_{\text{total}}$. In this respect, several principles are considered:
\begin{itemize}
\item Given that a single-shot (Stackelberg) negotiation protocol would inherently bias the outcome toward the leader agent, we adopt a constrained, multi-round protocol. Agent 1 (eMBB) and Agent 2 (URLLC) initiate the process by drawing an optimal starting baseline from a shared interaction Memory. Agents evaluate opponents' proposals against their strict constraints and retain the right to issue counter-proposals. This alternating exchange continues until a consensus is reached or a maximum round limit is triggered.
\item To ensure near-real-time execution on local 1B-parameter language models, the LLM component is strictly utilized for strategic goal-setting (generating exactly one optimal counter-proposal), while a deterministic 2D Proportional-Integral-Derivative (PID) algorithm conducts the fine-grained local search to mathematically guarantee feasibility.
\item It is imperative to distinguish between the sequential nature of the \textit{negotiation logic} and the concurrent nature of \textit{resource utilization}. While the decision-making process is iterative to strictly resolve conflict, both slices access and utilize the allocated shared resources simultaneously during the operational time slot $t$.
\end{itemize}

\section{Uncertainty and Risk-Aware Agentic System}
\label{sec:math_framework}

In this dynamic system, an autonomous agent $i$ must select an action $a_i = (b_i, f_i)$ from a set of possible actions $\mathcal{A}_i$. The outcome of this action is the uncertain E2E latency, $L_i(a_i) = L^{\text{edge}}_i(f_i) + L^{\text{RAN}}_i(b_i)$. The agent's objective is to satisfy its Service Level Agreement (SLA), $L_{i, \text{SLA}}$. Before making a proposal, the agent uses its DT to run a vectorized Monte Carlo simulation by sampling $SE_{i,t}$ to predict a full \textit{distribution} of potential latencies for the next timestep.

\subsection{Uncertainty and Risk Definition}
The DT takes a proposed action $a_i = (b_i, f_i)$ and the current state to generate a latency distribution $L_i(a_i)$ from $N_{mc}$ samples, i.e.,
\begin{equation}
\{L^{(k)}_i(a_i)\}_{k=1}^{N_{mc}} \text{ where } L^{(k)}_i(a_i) = L^{(e)}_i(f_i) + L^{(r, k)}_i(b_i).
\end{equation}
Here, $L^{(e)}_i(f_i)$ is the compute latency (derived from Eq. \ref{eq: comp_lat}) and $L^{(r, k)}_i(b_i)$ is the stochastic radio latency (derived from Eq. \ref{eq: radio_lat} using the $k$-th sample $SE^{(k)}_{i,t+1}$ from $\mathcal{U}[SE_{min}, SE_{max}]$). This output distribution allows us to quantify two distinct types of uncertainty as depicted in Figure \ref{fig: concept}.

\subsubsection{\textbf{Aleatoric Uncertainty (Statistical Risk)}}
This is the inherent randomness in the system, represented by the \emph{shape} of the distribution $L_i(a_i)$. We quantify its tail risk using the CVaR at an $\alpha$ confidence level (in our case, $\alpha=0.99999$). $\text{CVaR}_{\alpha}$ represents the expected latency in the worst $(1-\alpha)\%$ of cases, i.e.,
\begin{equation}
\text{VaR}_{\alpha}(L_i(a_i)) = \inf \{ l \in \mathbb{R} : P(L_i(a_i) \le l) \ge \alpha \},
\end{equation}
\begin{equation}
\text{CVaR}_{\alpha}(L_i(a_i)) = \mathbb{E}[L_i(a_i) | L_i(a_i) > \text{VaR}_{\alpha}(L_i(a_i))].
\end{equation}
In practice, VaR and CVaR are estimated empirically from the latency samples by sorting the samples and averaging those beyond the empirical $\alpha$-quantile.

\subsubsection{\textbf{Epistemic Uncertainty (DT Prediction Confidence)}}
This is the DT's \emph{uncertainty about its own prediction.} It measures how reliable the entire predicted distribution is. We model this using the coefficient of variation (CV) of the latency distribution and define the \emph{Epistemic Confidence Score} $C_E(a_i)$ as,
\begin{equation}
C_E(a_i) = \max \left( 0, 1 - \frac{\sigma_L(a_i)}{\mu_L(a_i)} \right),
\label{eq: conf}
\end{equation}
where $\mu_L(a_i)$ and $\sigma_L(a_i)$ are the mean and standard deviation of $L_i(a_i)$. A $C_E(a_i)$ score near 1.0 implies high confidence.

\subsection{Approach 1: Biased Agent (Neglect of Uncertainty)}
The biased agent exhibits two forms of cognitive bias: it \emph{neglects aleatoric risk} by performing an act of willful information reduction, discarding the rich data of the distribution's tail and basing its decisions solely on the mean $\mu_L(a_i)$; and it \emph{neglects epistemic risk} by ignoring the confidence score $C_E(a_i)$.

The agent's decision metric is therefore the simple expected latency,
\begin{equation}
\bar{L}_{\text{biased}}(a_i) = \mu_L(a_i) = \mathbb{E}[L_i(a_i)].
\label{eq: mean}
\end{equation}
The agent's final state of \emph{satisfaction} ($\mathcal{S}$) is governed by a brittle logical policy. It is based on only two conditions: is the SLA met ($\mathcal{M}$) based on the mean, and is the agent \emph{not} over-provisioning ($\mathcal{O}$)?
\begin{equation}
\mathcal{M}_{\text{biased}} \iff \bar{L}_{\text{biased}}(a_i) \le L_{i, \text{SLA}}
\end{equation}
\begin{equation}
\mathcal{O}_{\text{biased}} \iff \bar{L}_{\text{biased}}(a_i) < \theta_{\text{biased}} \cdot L_{i, \text{SLA}}
\end{equation}
where $\theta_{\text{biased}}$ is an aggressive threshold (e.g., $0.95$). The agent's final state of satisfaction is defined by the logical AND operation,
\begin{equation}
\mathcal{S}_{\text{biased}} \iff \mathcal{M}_{\text{biased}} \land \neg \mathcal{O}_{\text{biased}}.
\end{equation}
This policy is what defines the \emph{uncertainty neglect bias}. A low mean $\mu_L(a_i)$ can easily mask a high $\text{CVaR}_{\alpha}(a_i)$ that actually violates the SLA in practice.

\begin{algorithm}[t]
\SetAlgoLined
\DontPrintSemicolon
\footnotesize
\caption{Agentic Negotiation with 2D PID Search and Epistemic Risk Mitigation}
\label{alg:risk_aware_negotiation}

\KwIn{Agents $\mathcal{I}=\{1,2\}$, SLAs $\{L_{i,\text{SLA}}\}$, Constraints $\mathcal{C}_{\text{tot}} = \{B_{\text{tot}}, F_{\text{tot}}\}$, Max Rounds $N_{\text{rounds}}$}
\KwOut{Final Resource Allocation $\mathcal{A}^* = \{a_1^*, a_2^*\}$}

\textbf{Initialize:} Retrieve historical anchors $a_1, a_2$ from Memory $\mathcal{H}$\;

\For{$r \leftarrow 1$ \KwTo $N_{\text{rounds}}$}{
    \tcp{--- Phase 1: Risk-Aware Validation ---}
    \For{$i \in \mathcal{I}$}{
        $\{L^{(k)}_i\} \leftarrow \text{DT}(a_i)$ \tcp*{Vectorized Monte Carlo}
        Calculate $\mu_i$, $\sigma_i$, and $\text{CVaR}_{\alpha, i}$\;
        $C_E(a_i) \leftarrow \max(0, 1 - \sigma_i/\mu_i)$ \tcp*{Epistemic Conf.}

        \tcp{Mitigate Bias: Inflate Risk \& Apply Guardband}
        $\widetilde{\text{CVaR}}_{\alpha, i} \leftarrow \text{CVaR}_{\alpha, i} / \max(0.1, C_E(a_i))$\;
        $L^{gb}_{i,\text{SLA}} \leftarrow 0.8 \cdot L_{i,\text{SLA}}$\;
        
        $\mathcal{M}_i \leftarrow (\widetilde{\text{CVaR}}_{\alpha, i} \le L^{gb}_{i,\text{SLA}})$ \tcp*{SLA Met}
        $\mathcal{O}_i \leftarrow (\widetilde{\text{CVaR}}_{\alpha, i} < 0.7 \cdot L^{gb}_{i,\text{SLA}})$ \tcp*{Over-prov}
        $\var{Sat}_i \leftarrow \mathcal{M}_i \land \neg \mathcal{O}_i$\;
    }

    \If{$\var{Sat}_1 \land \var{Sat}_2$ \textbf{and} $\sum a_i \le \mathcal{C}_{\text{tot}}$}{
        $\mathcal{H}.\text{update}(a_1, a_2, \text{success})$\;
        \KwRet{$\{a_1, a_2\}$} \tcp*{Consensus Reached}
    }

    \tcp{--- Phase 2: LLM Proposal \& 2D PID Search ---}
    \For{$i \in \mathcal{I}$}{
        $j \leftarrow \mathcal{I} \setminus \{i\}$ \tcp*{Opponent}
        $\mathcal{C}_{\text{rem}} \leftarrow \{B_{\text{tot}}-b_j, F_{\text{tot}}-f_j\}$\;
        
        $\var{Prompt} \leftarrow \textrm{ConstructContext}(a_j, \mathcal{C}_{\text{rem}}, \widetilde{\text{CVaR}}_{\alpha, i})$\;
        $a^*_i \leftarrow \textrm{LLM\_API}(\var{Prompt})$ \tcp*{Generate 1 target}
        
        \tcp{Deterministic 2D Proportional Correction}
        \While{$\neg \var{Sat}_i(a^*_i)$ within limits of $\mathcal{C}_{\text{rem}}$}{
            Calculate severity error $e_i \leftarrow \widetilde{\text{CVaR}}_{\alpha, i} / L^{gb}_{i,\text{SLA}}$\;
            Determine adaptive steps $\Delta b, \Delta f \propto e_i$\;
            Identify bottleneck queue (Radio vs. Compute)\;
            Adjust $b^*_i, f^*_i$ proportionally to resolve bottleneck or optimize energy\;
        }
        $a_i \leftarrow a^*_i$\;
    }
}
$\mathcal{H}.\text{update}(a_1, a_2, \text{failure})$\;
\KwRet{$\{a_1, a_2\}$} \tcp*{Return final state}
\end{algorithm}

\subsection{Approach 2: Unbiased Agent (CVaR-Aware Mitigation)}
The unbiased agent is designed to actively mitigate both forms of risk simultaneously by combining a \emph{Greedy Guardband} buffer with an \emph{Uncertainty-Adjusted Risk} metric.

First, to absorb extreme tail fluctuations, the agent targets a strict Safety Guardband, defined as a percentage $\beta$ of the true SLA,
\begin{equation}
L^{gb}_{i, \text{SLA}} = \beta \cdot L_{i, \text{SLA}}, \quad (\text{e.g., } \beta = 0.8).
\end{equation}
Second, it mitigates epistemic risk by incorporating the Confidence Score $C_E(a_i)$ directly into the risk evaluation. If the DT detects high variance (low confidence), it artificially inflates the measured aleatoric risk, yielding the \emph{Uncertainty-Adjusted Risk} $\widetilde{\text{CVaR}}_{\alpha}$,
\begin{equation}
\widetilde{\text{CVaR}}_{\alpha}(a_i) =
\begin{cases}
\text{CVaR}_{\alpha}(a_i), & \text{if } C_E(a_i) \ge 0.9 \\
\frac{\text{CVaR}_{\alpha}(a_i)}{\max(0.1, C_E(a_i))}, & \text{otherwise}
\end{cases}.
\label{eq: adjusted_cvar}
\end{equation}
The agent's core success check $\mathcal{M}_{\text{unbiased}}$ compares this strictly inflated risk against the guardband target,
\begin{equation}
\mathcal{M}_{\text{unbiased}}(a_i) \iff \widetilde{\text{CVaR}}_{\alpha}(a_i) \le L^{gb}_{i, \text{SLA}}.
\end{equation}
Furthermore, to ensure strict SLAs like URLLC are never compromised by premature energy optimization, the agent adopts a \emph{Greedy Resource Holding} mechanism. It refuses to downward-optimize resources unless it is massively over-provisioned,
\begin{equation}
\mathcal{O}_{\text{unbiased}} \iff \widetilde{\text{CVaR}}_{\alpha}(a_i) < \theta_{\text{unbiased}} \cdot L^{gb}_{i, \text{SLA}},
\end{equation}
where $\theta_{\text{unbiased}}$ enforces a highly conservative threshold (e.g., $0.7$). The final satisfaction state remains,
\begin{equation}
\mathcal{S}_{\text{unbiased}} \iff \mathcal{M}_{\text{unbiased}} \land \neg \mathcal{O}_{\text{unbiased}}.
\label{eq: satisf}
\end{equation}
The integration of this risk-aware utility function with the 2D PID search algorithm is detailed in Algorithm \ref{alg:risk_aware_negotiation}.

\subsection{Meta-Verification in Practice: Propagating Confidence}
The \texttt{confidence\_score} derived from Eq. \ref{eq: conf} serves as the practical engine for meta-verification. Its utility is realized by actively propagating this trust metric through the agent's entire decision-making pipeline.

\textbf{(i) Phase 1: Internal Risk-Aware Validation,} where the agent performs an internal self-assessment. Its Digital Twin simulates the current state using vectorized Monte Carlo sampling, integrating the empirical tail risk (CVaR) and the \texttt{confidence\_score} to compute the Uncertainty-Adjusted Risk ($\widetilde{\text{CVaR}}_{\alpha}$). This mechanism mathematically penalizes low-confidence predictions, forcing the agent to evaluate its true performance limits against a strict Greedy Guardband target ($L^{gb}_{i, \text{SLA}}$).

\textbf{(ii) Phase 2: Generative Goal-Setting and Proportional Correction,} where the agent leverages a local, fast-inference 1B-parameter model (\texttt{otel-llm-1b-it}) to generate a single strategic target proposal based on the validated risk metrics. This generative target is then iteratively refined by a deterministic 2D Proportional-Integral-Derivative (PID) algorithm, which adjusts bandwidth and CPU allocations proportionally to the bottleneck severity until the strict SLA and feasibility constraints are mathematically guaranteed. This hybrid approach ensures both transparent reasoning and near-real-time execution.

This act of rigorously evaluating and reporting confidence elevates the negotiation from a simple exchange of demands to a sophisticated, verifiable dialogue about uncertainty.

\section{Experimental Results}
\label{sec:results}

To validate the theoretical framework, we conducted a series of simulations comparing the performance of the two agent strategies.

\subsection{Simulation Setup}
The simulation environment is configured with two slice agents, namely, eMBB ($L_{1,\text{SLA}} = 50\text{ ms}$) and URLLC ($L_{2,\text{SLA}} = 10\text{ ms}$). To ensure near-real-time execution and data privacy, the agents utilize a locally deployed, 1-billion-parameter language model (\texttt{otel-llm-1b-it}) executed on a single NVIDIA RTX A4000 GPU for fast strategic inference, accessed via an OpenAI-compatible local API. The agents negotiate over a total RAN bandwidth of $B_{\text{total}} = 100.0\text{ MHz}$ and a total edge compute capacity of $F_{\text{total}} = 100.0\text{ GHz}$. The edge compute processing rate is defined as $R_{\text{cpu}} = 20\text{ Mbps per GHz}$.

Each simulation comprises $N_{\text{trials}} = 200$ independent negotiations, with each negotiation restricted to up to $N_{\text{rounds}} = 5$ rounds. The base traffic arrival rates are set to $90\text{ Mbps}$ for eMBB and $40\text{ Mbps}$ for URLLC, governed by time-varying burst intervals. The main sources of aleatoric uncertainty arise from the Spectral Efficiency (SE), which fluctuates stochastically within the range $\mathcal{U}[5.0, 7.0]$, alongside the epistemic confidence of the DT predictions. The power consumption model is defined by a static base power of $P_{\text{static}} = 5.0\text{ W}$, a bandwidth power cost of $C_{\text{bw}} = 0.5\text{ W/MHz}$, and a linear CPU power cost of $C_{\text{CPU}} = 0.001\text{ W/GHz}^3$.

Two agent strategies are compared: (1) a Biased (Neglect of Uncertainty) strategy, where agents rely solely on the mean latency $\mu_L(a_i)$ as defined in Eq.~(\ref{eq: mean}), and (2) an Unbiased (Risk-Aware Mitigation) strategy, where decisions are governed by the Uncertainty-Adjusted Conditional Value-at-Risk $\widetilde{\text{CVaR}}_{\alpha}$, the Epistemic Confidence $C_E$, and the Greedy Guardband limits, as formulated in Eqs.~(\ref{eq: adjusted_cvar})--(\ref{eq: satisf}).

\begin{figure}[t]
    \centering
\includegraphics[width=0.45\textwidth]{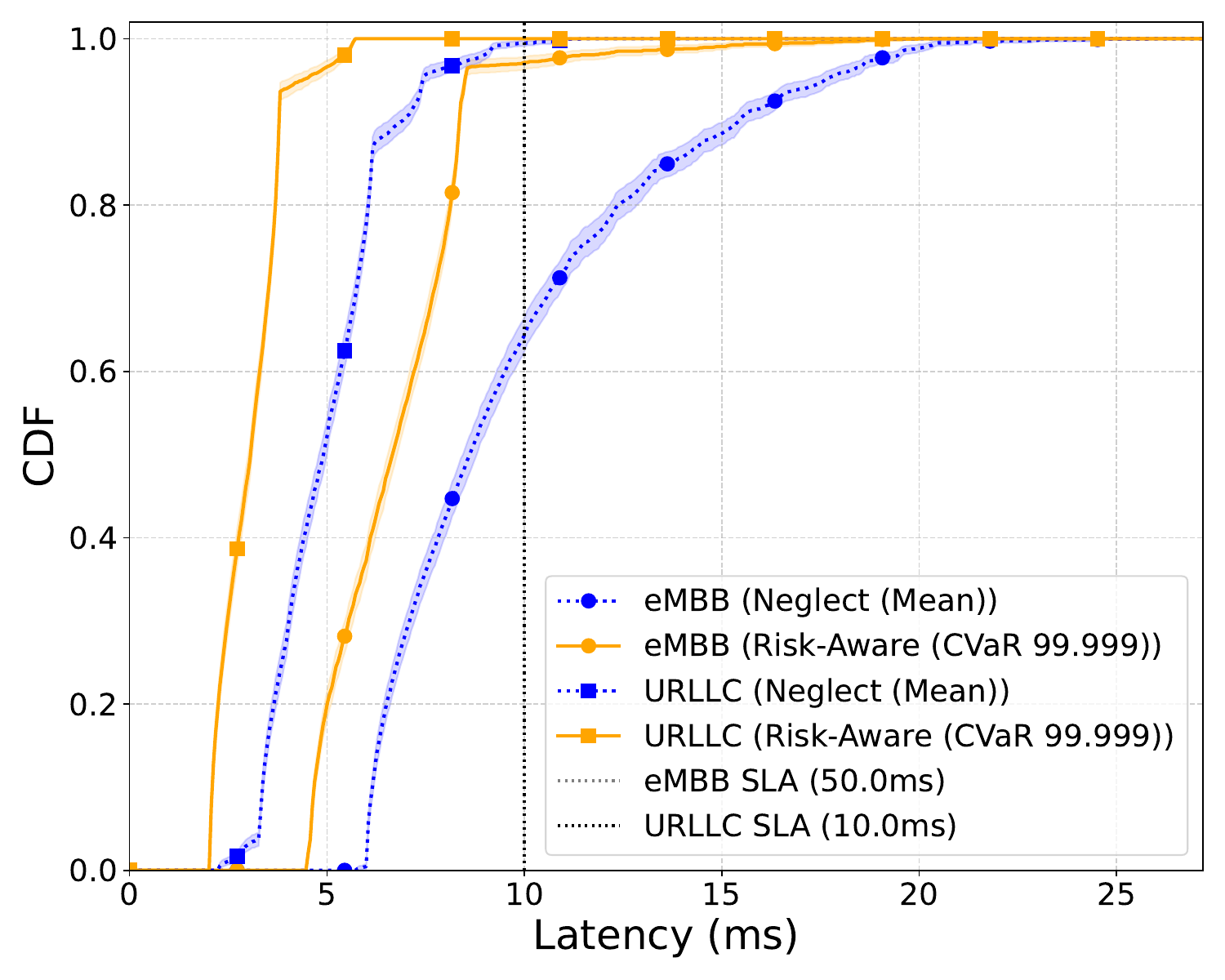}
    \caption{Latency CDF for both agents vs. various scenarios.}
    \label{fig:latency_cdf}
\end{figure}

\subsection{Analysis of Service Level Agreement (SLA) Performance}
\begin{figure}[t]
  \centering
  \subfloat[Total SLA violations for various scenarios.]{%
 \includegraphics[width=0.20\textwidth]{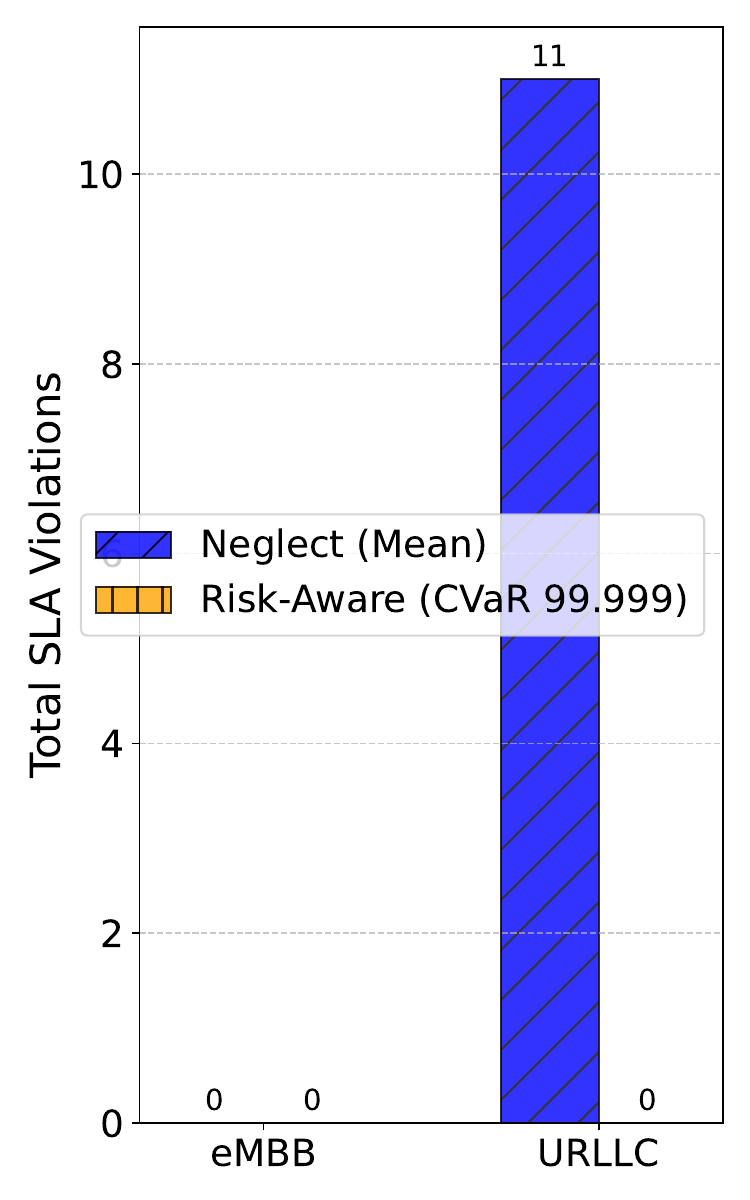}
    \label{fig:age}
  }
  \hfill
  \subfloat[The 99.999th-percentile (p99.999) latency for various scenarios.]{%
 \includegraphics[width=0.20\textwidth]{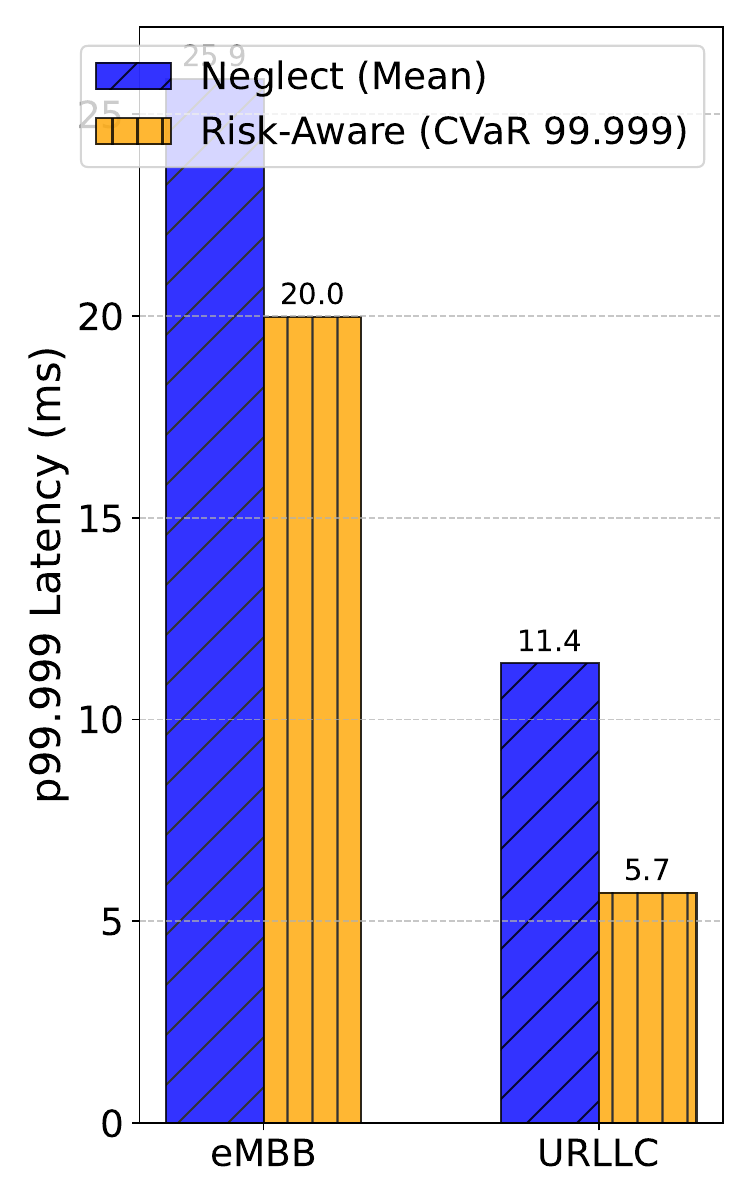}
    \label{fig:ratio}
  }
  \caption{Debiasing gain from the risk-Aware (CVaR) strategy vs. the biased mean-based reasoning.}
  \label{fig:gain}
\end{figure}

The primary goal of the agent is to satisfy its SLA. Figure~\ref{fig:latency_cdf} and Figure~\ref{fig:gain} provide a comprehensive view of SLA performance. As shown in Figure~\ref{fig:latency_cdf}, the choice of strategy has a dramatic impact on SLA compliance. \textbf{URLLC (Strict 10ms SLA):} The Biased agent (blue dotted line with squares) fails systematically. Its latency CDF reaches up to approximately 15ms, heavily violating the strict deadline. In contrast, the Unbiased (Risk-Aware) agent (orange solid line with squares) successfully optimizes for the tail, keeping its entire latency distribution bounded comfortably below 10ms (reaching unity around 6ms). \textbf{eMBB (Relaxed 50ms SLA):} While the 50ms SLA gives eMBB more headroom, the Unbiased agent remains much more conservative, tightening the latency distribution significantly compared to the Biased agent.

Figure~\ref{fig:gain} quantifies the practical impact of mitigating the uncertainty neglect bias. \textbf{Total SLA Violations (Left):} The bar chart reveals that the Biased agent caused 11 critical SLA violations for the URLLC slice across the 200 trials. This confirms that the tail-risk observed in the CDF translates directly to service failure. Conversely, the Unbiased agent successfully mitigated all uncertainty, registering exactly 0 violations for both slices. \textbf{p99.999 Latency (Right):} This metric exposes the hidden vulnerability of the mean-based approach. The Unbiased agent achieves a massive reduction in tail risk. For eMBB, it cuts the p99.999 latency by 51.7\% (from 17.6ms down to 8.5ms). Most critically, for URLLC, it reduces the p99.999 latency from 7.4ms down to 4.4ms. While the 7.4ms p99.999 value for the Neglect scenario seems safe, its remaining 5\% tail is what caused the 11 catastrophic violations, emphasizing the absolute necessity of reasoning over the strict $\widetilde{\text{CVaR}}_{\alpha}$ metric to guarantee safety.

\subsection{Analysis of Energy Saving}
\begin{figure}[t]
    \centering
 \includegraphics[width=0.45\textwidth]{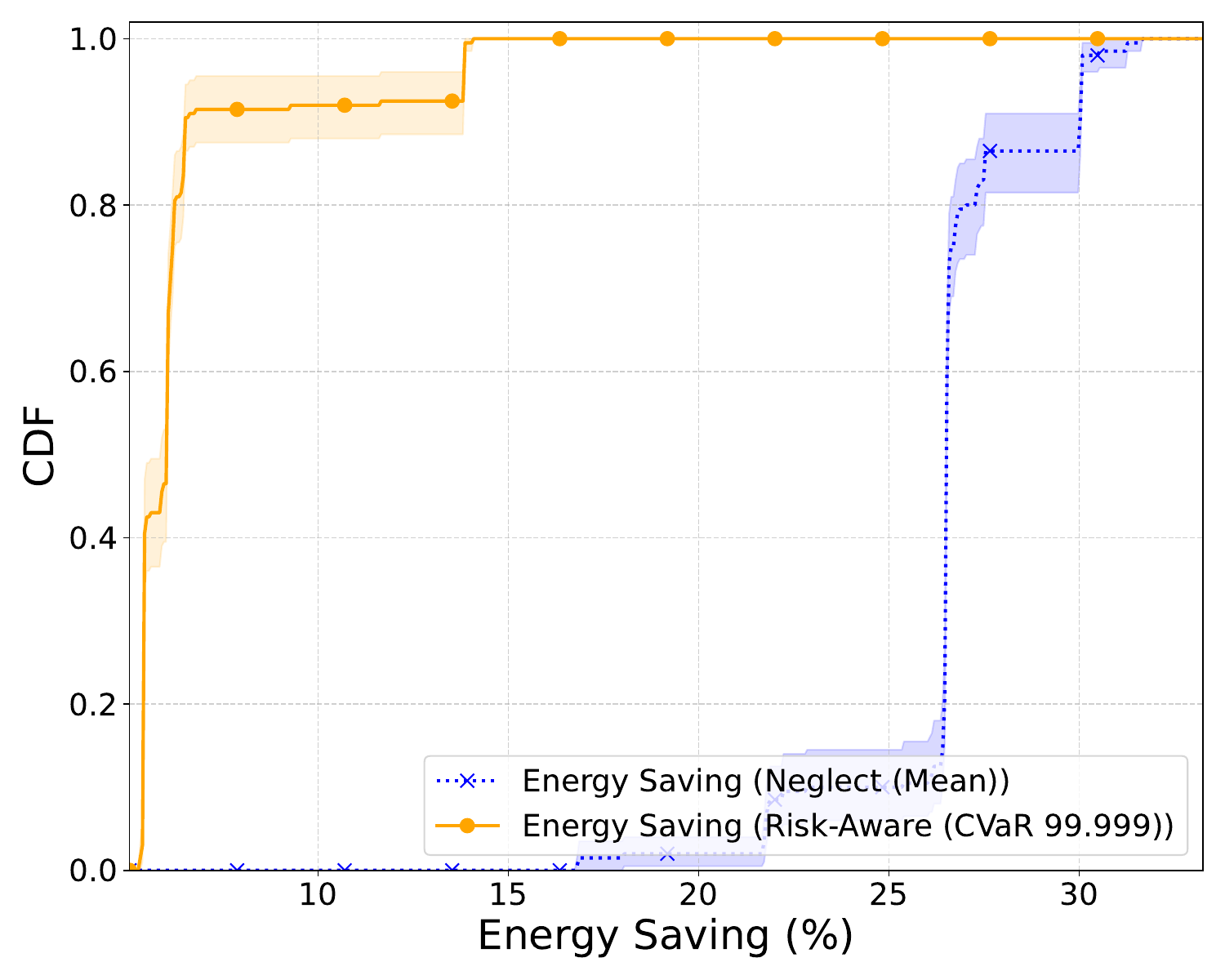}
    \caption{CDF of Energy Saving for both slices vs. scenarios.}
    \label{fig:energy_cdf}
\end{figure}

SLA compliance is not the only objective; agents also attempt to minimize resource footprints to yield energy savings. Figure~\ref{fig:energy_cdf} reveals the empirical trade-off inherent in the two strategies. The Energy Saving CDF for the \textbf{Unbiased (Risk-Aware)} agent (orange solid line) is visibly shifted to the left of the \textbf{Biased (Neglect)} agent (blue dotted line). The risk-aware agent consistently achieves lower energy savings, predominantly clustered between 2\% and 15\%. In stark contrast, the biased agent aggressively pursues savings, clustering closely between 26\% and 30\%. 

This divergence is not a failure of the risk-aware model; it is the mathematically rational cost of reliability. The unbiased agent actively buys its robust SLA compliance (Figure~\ref{fig:latency_cdf}) by refusing to over-optimize. It purposefully holds a statistically grounded bandwidth and CPU buffer to protect against tail risk fluctuations and epistemic uncertainty. The biased agent operates on a \emph{false economy}: it appears more energy-efficient on average, but achieves this exclusively by operating dangerously close to the failure boundary, remaining entirely blind to the stochastic spikes that ultimately cause it to fail its primary operational objective.

\begin{figure}[t]
    \centering
 \includegraphics[width=0.45\textwidth]{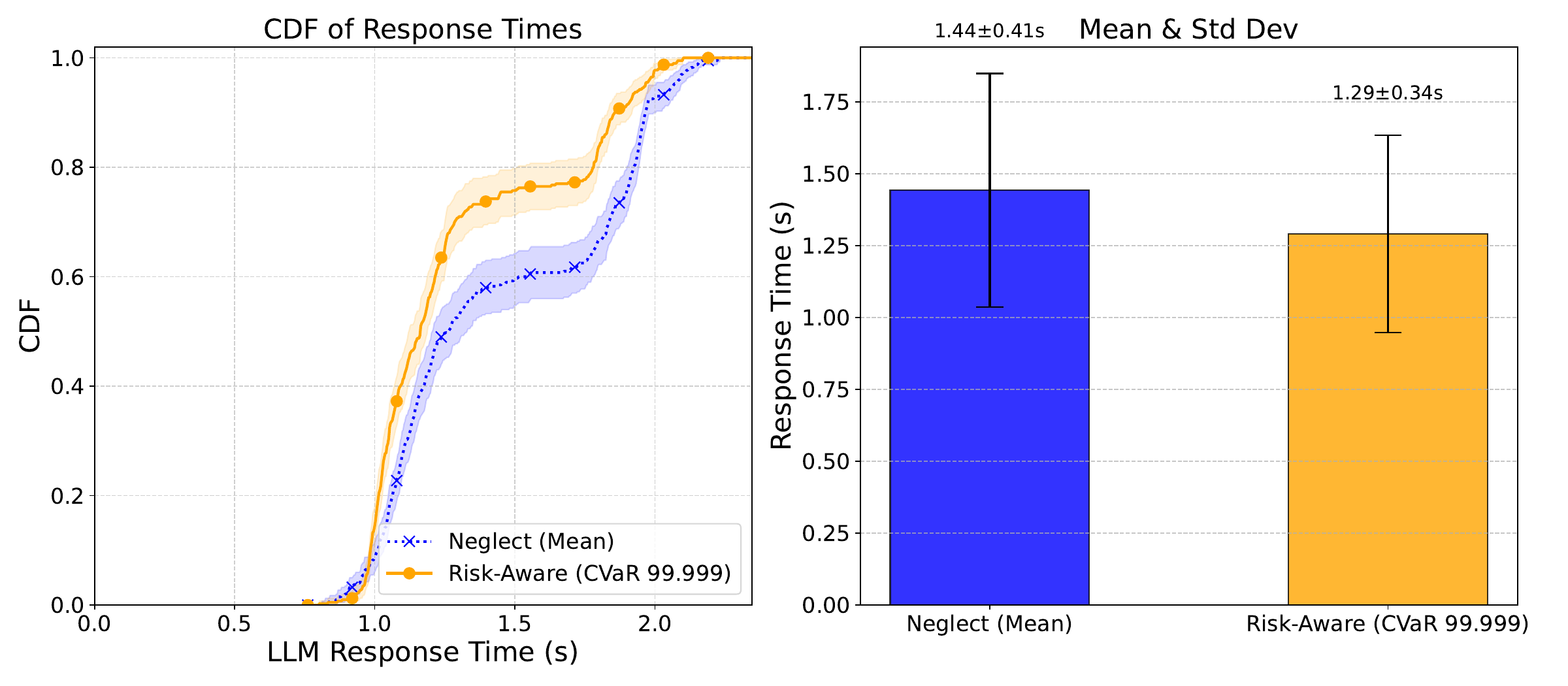}
    \caption{LLM inference time using a local 1B model on an RTX A4000 GPU.}
    \label{fig:llm_resp_time}
\end{figure}

\subsection{LLM Inference Time and Real-Time Feasibility}
To validate the practical deployability of our framework for 6G operational speeds, we assessed the LLM response times. The negotiations were powered by the local \texttt{otel-llm-1b-it} model executed on a single NVIDIA RTX A4000 GPU. Figure~\ref{fig:llm_resp_time} illustrates the CDF and statistical distribution of these inference times.

The system consistently achieves sub-two-second inferences. The Biased (Neglect) strategy required a mean response time of 1.44s ($\pm$ 0.41s), while the Risk-Aware (CVaR) strategy slightly outperformed it with a mean of 1.29s ($\pm$ 0.34s). This performance confirms that edge-deployed, 1-billion-parameter LLMs possess the computational efficiency required to perform advanced risk reasoning and goal-setting dynamically. By offloading the fine-grained numerical convergence to the deterministic 2D PID algorithm and restricting the LLM to high-level strategic single-shot proposals, the framework avoids catastrophic execution delays. This proves the viability of non-real-time RIC-aligned agentic negotiation without reliance on slow, privacy-compromising cloud APIs.

\section{Conclusion}
\label{sec:conclusion}
Trustworthy 6G autonomous networks require agentic systems that can reason beyond simple averages and account for high-stakes, low-probability events. This paper addressed the critical \emph{uncertainty neglect bias} through proposing a risk-aware negotiation framework. By compelling agents to use Conditional Value-at-Risk for reasoning over latency tails and DTs to quantify and propagate their own \emph{epistemic} uncertainty, our approach ensures robust reasoning over resource allocation. We demonstrated in a 6G cross-domain inter-slice negotiation use case that this unbiased, risk-aware method effectively eliminates the SLA violations endemic to mean-based approaches, significantly reduces p99.999 latencies, and rationally quantifies the trade-off between reliability and efficiency. Furthermore, executing this logic via an edge-hosted 1B-parameter model on a single RTX A4000 GPU achieves inference times of $\sim$1.3 seconds, providing a concrete, low-latency methodology for building the robust, verifiable, and trustworthy autonomous agents required for future 6G systems.

\bibliographystyle{IEEEtran}
\bibliography{bibliography.bib}

\end{document}